\documentclass{PoS}

\title{Energy dependent forward B $\rightarrow$ $J/\psi$ measurements in p+p collisions at PHENIX}

\ShortTitle{Energy dependent forward B $\rightarrow$ $J/\psi$ measurements in p+p collisions at PHENIX}

\author{\speaker{Xuan Li (for the PHENIX Collaboration)} \\ 
        P-25, MS H846, Physics Divison, LANL, Los Alamos, NM, 87545, USA\\
        E-mail: \email{xuanlipx@rcf.rhic.bnl.gov}}

\abstract{The heavy flavor studies at RHIC help improve the knowledge of the bottom/charm quark production and can test Quantum Chromodynamics (QCD). Compared to the LHC/Tevatron, the RHIC heavy flavor production originates from different partonic sub-processes and has a complementary kinematic coverage. The PHENIX forward rapidity silicon vertex detector (FVTX) provides precise determination of the event vertex, tracking and the Distance of Closest Approach (DCA) of charged tracks. This detector allows direct access to the $B$ meson production via measurements of non-prompt $J/\psi \to \mu^{+} + \mu^{-}$ within $1.2<|y|<2.2$ rapidity in $p$+$p$ collisions at $\sqrt{s} = $ 510 and 200 GeV. Comparison among PHENIX measurements of the $B \rightarrow J/\psi$ fraction with integrated $J/\psi$ $p_{T}$ up to 5 GeV$/c$ and higher energy results at the Tevatron and the LHC presents a smooth center of mass energy dependence from 0.2 to 13 TeV in $p$+$p$ ($p$+$\bar{p}$) collisions. The Next-To-Leading order Perturbative QCD (NLO pQCD) calculations are in reasonable agreement with the extracted total $b\bar{b}$ cross section based on the $B \rightarrow J/\psi$ fraction measurements at PHENIX.}

\FullConference{XXV International Workshop on Deep-Inelastic Scattering and Related Subjects\\
		3-7 April 2017\\
		University of Birmingham, UK}

\begin{document}

\section{Introduction}
Heavy quarks are produced in the early stage of high energy hadronic collisions due to their high mass ($m_{\rm{c,b}} \gg \Lambda_{\rm{QCD}}$). Since the heavy flavor quarks do not vanish or change into other flavors during the hard scattering and the sequential fragmentation processes, they can be treated as hard probes to study the initial state partons. At RHIC energies, the bottom production is dominated by the gluon-gluon fusion (pair creation) process in contrast to the Tevatron and the LHC measurements which are dominated by the flavor excitation process \cite{Norrbin:2000zc}. In additional to this, the kinematic range of the nucleon (nuclear) gluon distribution function (PDF) accessed by the RHIC heavy flavor measurements is different from what has been measured at the LHC. Therefore, measuring the heavy flavor production at RHIC provides a unique test of Quantum Chromodynamics (QCD).

Yields of non-prompt $J/\psi$ from $B$ meson decay are proportional to the total $B$ meson production. Extracting $B$ meson production via this channel provides a better signal to background ratio compared to the bottom semi-leptonic decay measurements as backgrounds are suppressed by the $J/\psi$ identification. Decayed particles with a non-zero displaced vertex can be identified with the help of the Forward Silicon Vertex Detector (FVTX) at PHENIX, which can precisely measure the Distance of Closet Approach (DCA) of final state particles. Using this technique, $B$ mesons in the forward/backward rapidities have been studied at PHENIX in $p$+$p$ collisions. The $B \rightarrow J/\psi$ fractions have been measured and the $b\bar{b}$ cross sections are studied in 510 and 200 GeV $p$+$p$ collisions. The NLO pQCD calculations are in good agreements with these results.

\begin{figure}[h]
\centerline{\includegraphics[width=0.47\linewidth]{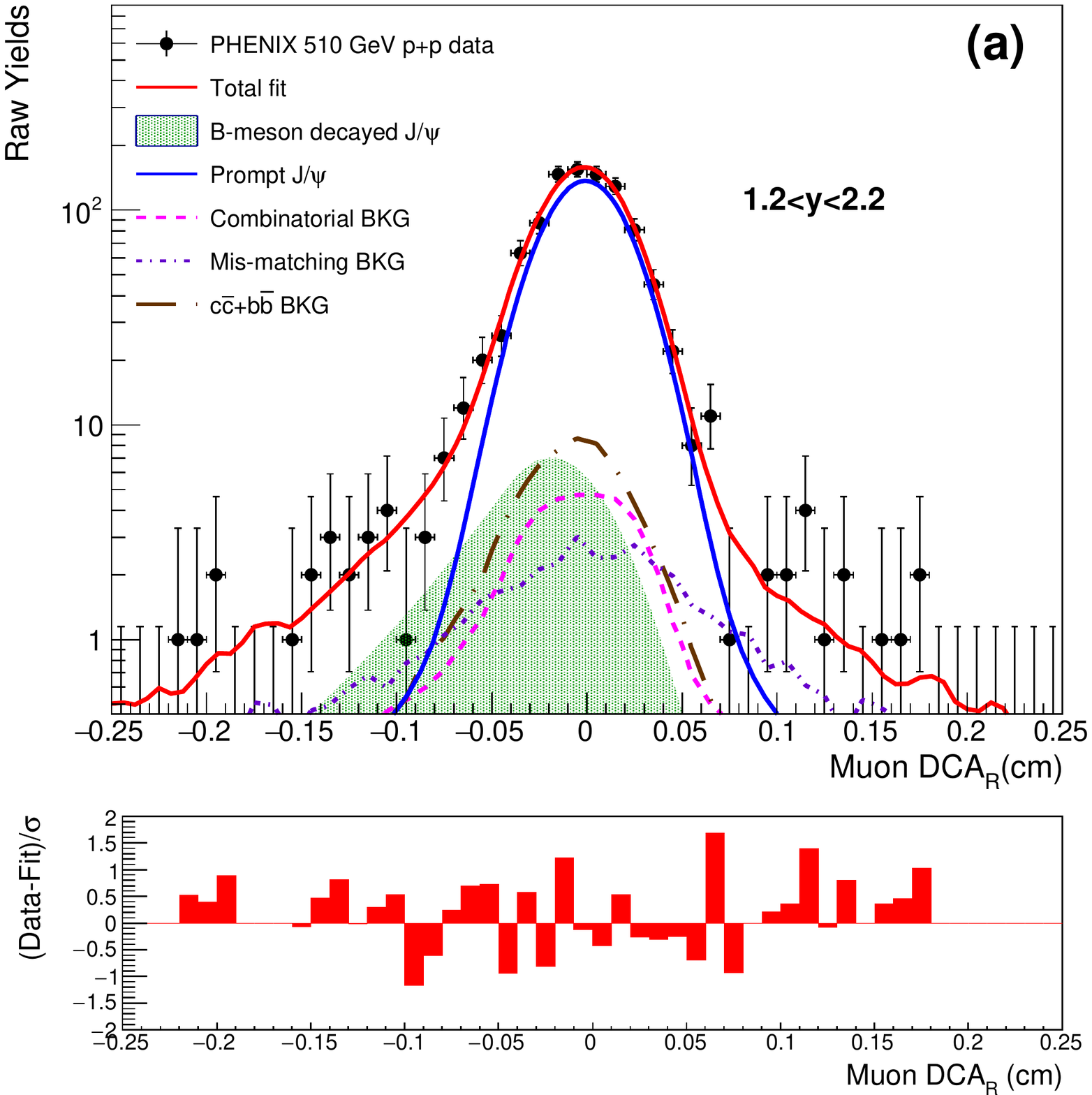}
\includegraphics[width=0.47\linewidth]{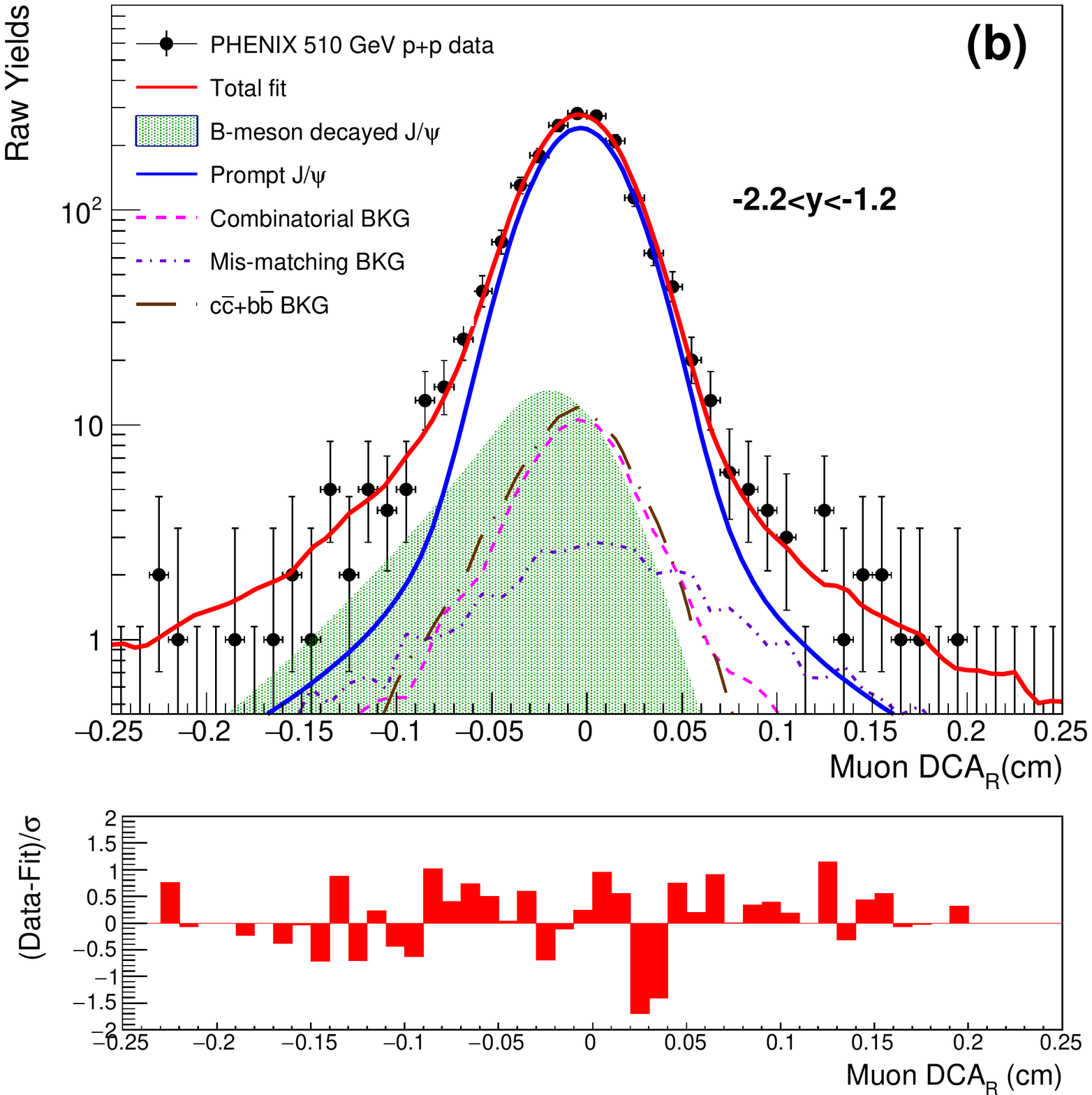}
}
\caption{ \it The $B \to J/\psi$ fraction fit to muon $\textrm{DCA}_{\rm R}$ of $J/\psi$ sample in the (a) $1.2<y<2.2$ and (b) $-2.2<y<-1.2$ regions. The red solid curve stands for the total fit which includes the prompt $J/\psi$ (solid blue), the $B$-meson $\rightarrow J/\psi$ (green filled region), the combinatorial background (magenta dashed curve), the $c\bar{c}+b\bar{b}$ background (brown long-dashed curve) and the detector mismatching background (purple short-dashed curve). Figures from \cite{ppg197}.}
\label{fig:dcar_fit}
\end{figure}

\section{The fraction of $J/\psi$ from $B$ meson decay in $\sqrt{s}$ = 510 and 200 GeV $p$+$p$ collisions}
The fraction of $J/\psi$ from $B$ meson decay at forward/backward rapidities ($1.2<|y|<2.2$) has been first studied at PHENIX in the 510 GeV $p$+$p$ collisions during the 2012 RHIC run. Identification of $J/\psi$ from $B$ meson decay is based on measurements of the DCA radial projection ($\textrm{DCA}_{\rm R}$) of muons from $J/\psi$ decay as the FVTX has a better spatial resolution along the radial direction compared to its azimuthal angle component. The $\textrm{DCA}_{\rm R}$ of muons from prompt $J/\psi$ decay follows a symmetric distribution. Because of the decay kinematics, muons from $B$ meson decayed $J/\psi$ produce an asymmetric tail in their $\textrm{DCA}_{\rm R}$ distribution.
 \begin{figure}[h]
 	\begin{center}
	\includegraphics[width=0.50\linewidth]{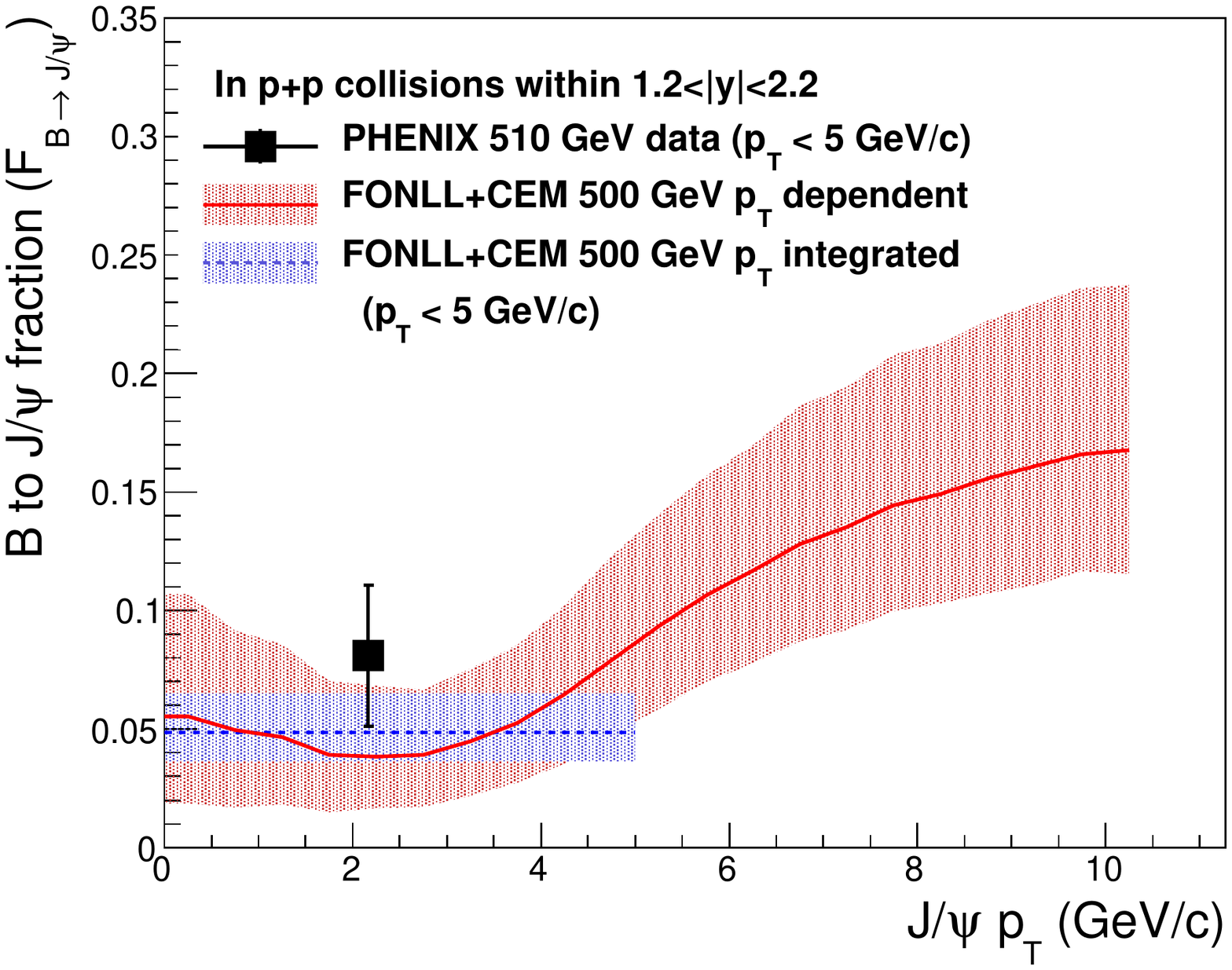}
	\includegraphics[width=0.47\linewidth]{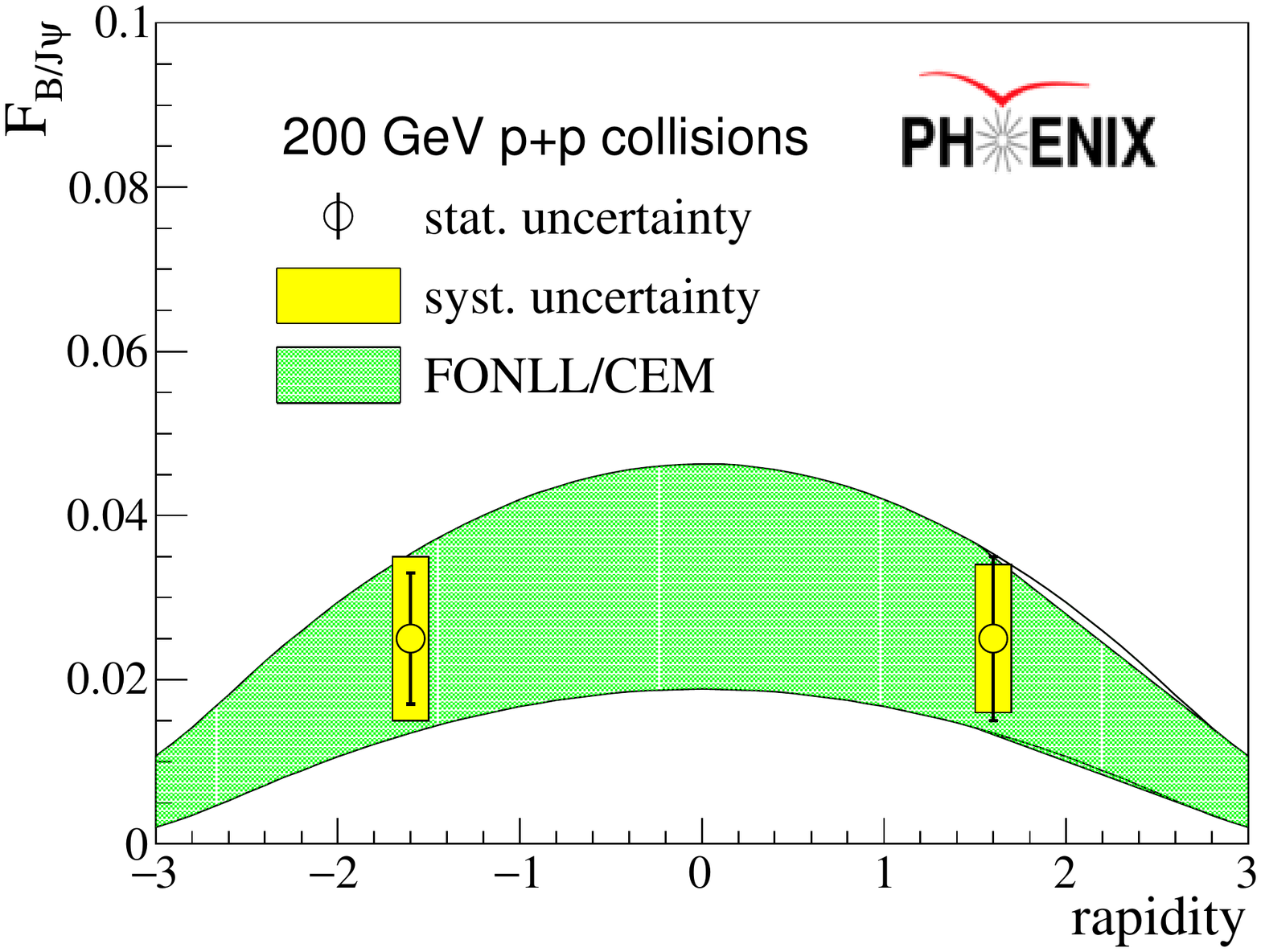}
	\caption{\label{fig:bfrac_pp} The $B \to J/\psi$ fraction with integrated $J/\psi$ $p_{T}$ and $1.2<|y|<2.2$ rapidity in $p$+$p$ collisions at $\sqrt{s} = $ 510 (left) GeV and $\sqrt{s} = $ 200 (right) GeV. In the left panel, the red shaded band stands for the FONLL+CEM calculated $p_{T}$ dependent $B \to J/\psi$ fraction within $1.2<|y|<2.2$ in 500 GeV $p$+$p$ collisions and the blue shaded curve represents the $p_{T}$ integrated ($p_{T}<$ 5 GeV$/c$)$ B \to J/\psi$ fraction within the same kinematic region. The yellow shaded curve in the right panel stands for the FONLL+CEM predicted rapidity dependent $B \to J/\psi$ fractions in 200 GeV $p$+$p$ collisions.}
\end{center}
\end{figure}

This analysis requires muon tracks pass through quality cuts and have good matching between the FVTX and the Muon Tracker (MuTr). $J/\psi$ candidates are selected from unlike-sign dimuon pairs within the mass region of 2.7-3.5 GeV/$c^{2}$. After verifications of consistent $\textrm{DCA}_{\rm R}$ resolutions in both data and the full simulation of PYTHIA+GEANT+reconstruction with realistic vertex and dead maps etc, the $\textrm{DCA}_{\rm R}$ distribution shapes of muons from prompt $J/\psi$ and $B$ meson decayed $J/\psi$ are determined in the full simulation. The detector mis-alignments have been corrected before and after the data production. Any remaining $\textrm{DCA}_{\rm R}$ offset is determined by identified prompt hadrons in data. Besides the prompt $J/\psi$ and $B \rightarrow J/\psi$ components, the dimuon sample in data also contains the combinatorial background, the detector mis-matching background and the heavy flavor ($c\bar{c}+b\bar{b}$) continuum background. The combinatorial background which represents the mis-identified muon and/or hadron pairs is determined by unlike-sign track pairs in normalized mixed event samples. Since a nearly 1 $m$ long hadron absorber is located between the FVTX and the MuTr, there is a certain probability of matching an incorrect FVTX track to a MuTr track after the multiple scattering within the absorber. The FVTX-MuTr mis-matching background is determined in mixed events with topologically rotated FVTX. The heavy flavor continuum background is determined using a full simulation. Details of the analysis procedure have been discussed in \cite{ppg197}. A maximum log-likelihood fit of the muon $\textrm{DCA}_{\rm R}$ distribution is applied to data to simultaneously determine the raw yields of prompt $J/\psi$ and non-prompt $J/\psi$ from $B$ meson decay. Figure \ref{fig:dcar_fit} shows the fit results within $-2.2<y<-1.2$ and $1.2<y<2.2$ rapidity regions in 510 GeV $p$+$p$ collisions.

\begin{figure}[t]
\centerline{\includegraphics[width=0.68\linewidth]{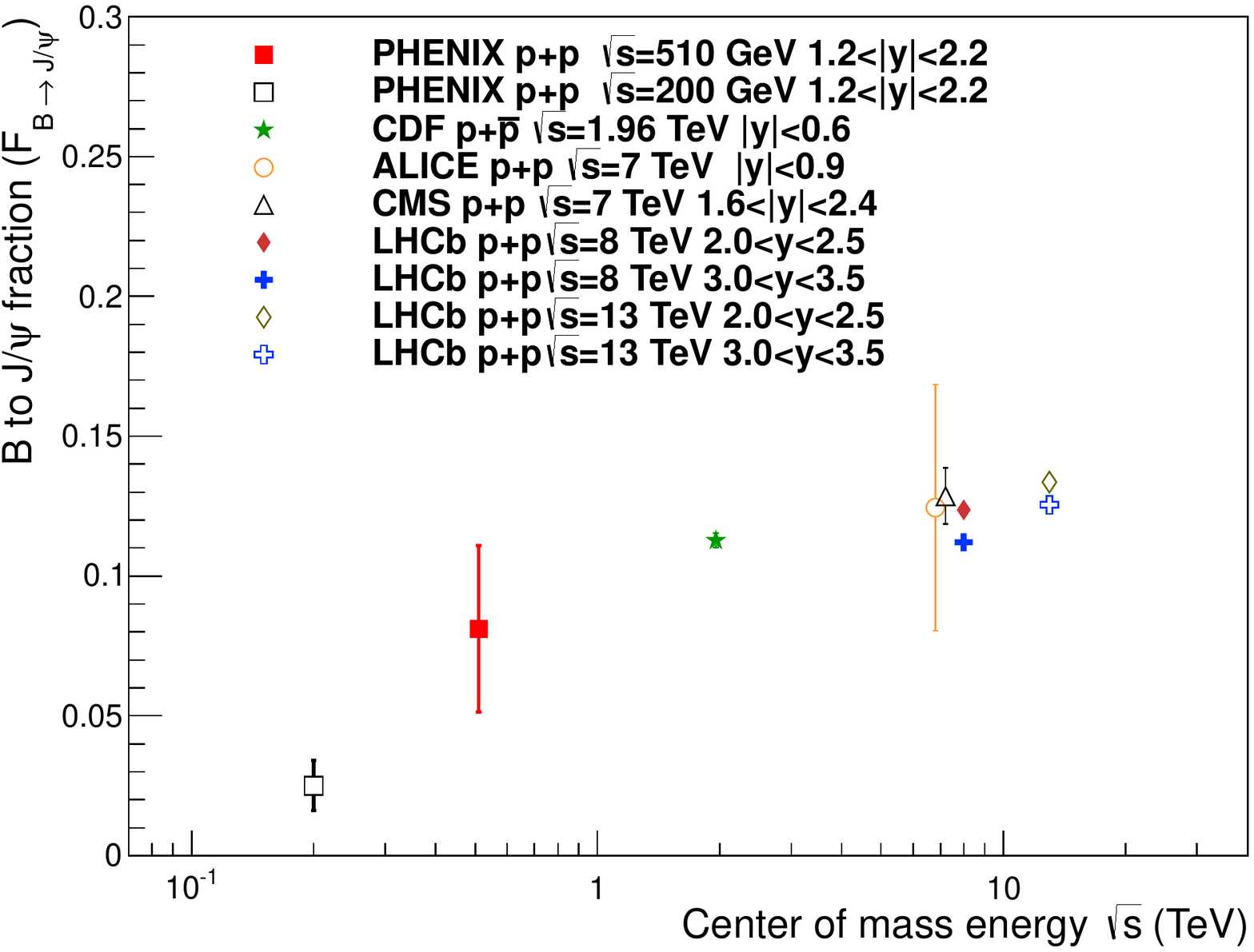}}
\caption{ \it Comparison of PHENIX $B$ $\rightarrow$ $J/\psi$ fraction measured in 510 GeV \cite{ppg197} and 200 GeV $p$+$p$ collisions \cite{ppg199} with the global data from CDF, ALICE, CMS and LHCb experiments as a function of center of mass energy integrated in the $J/\psi$ $0<p_{T}<5$ GeV/$c$ interval. The uncertainty is statistical and systematic combined.}
\label{fig:energy_btojpsi}
\end{figure}
\begin{figure}[h]
\centerline{\includegraphics[width=0.72\linewidth]{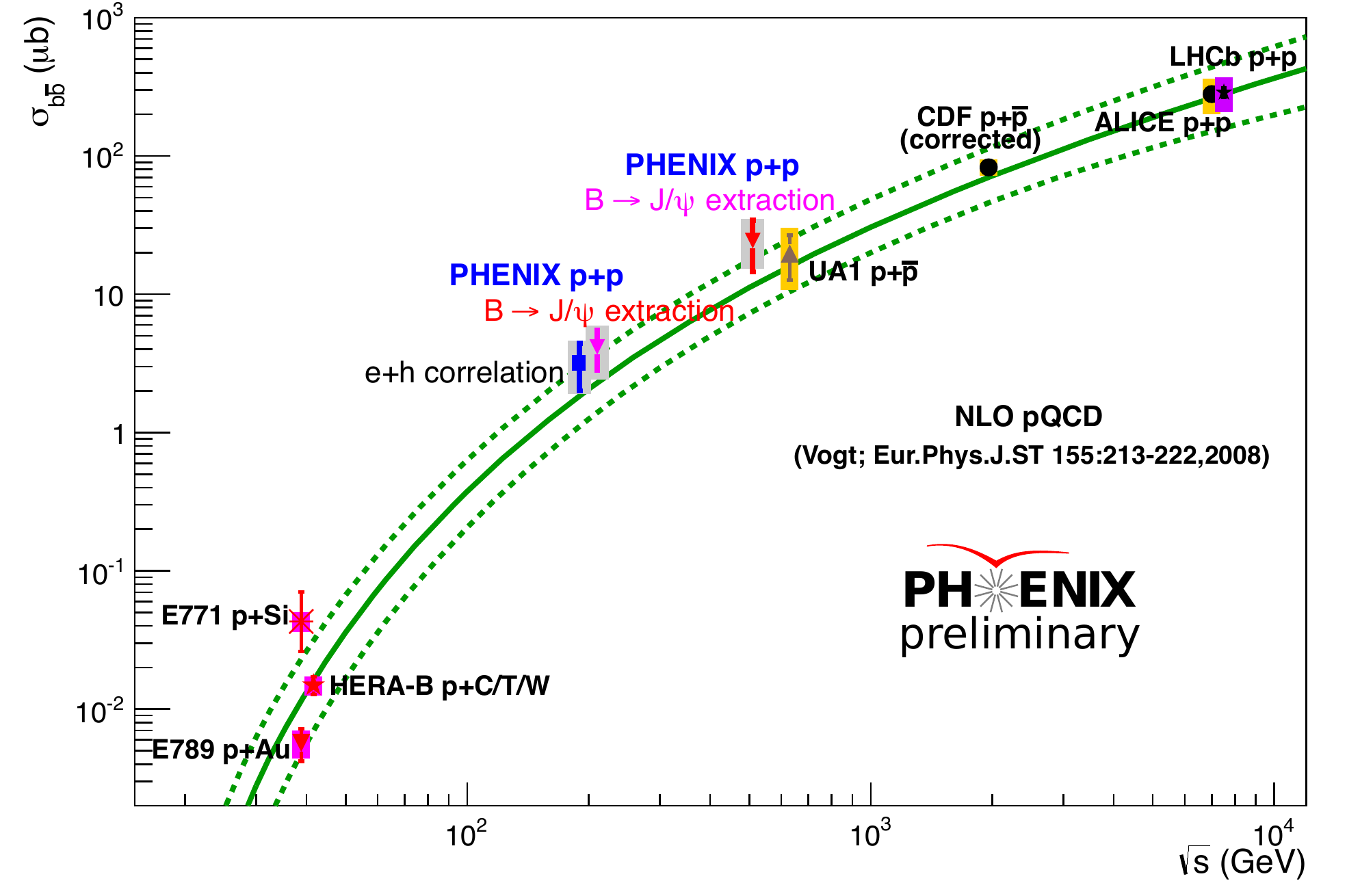}}
\caption{ \it Comparison of the $b\bar{b}$ cross section extracted from the PHENIX $B$ $\rightarrow$ $J/\psi$ fraction measured in 510 GeV \cite{ppg197} and 200 GeV \cite{ppg199} $p$+$p$ collisions with previous measurements at Fermilab, RHIC, LHC and NLO pQCD calculations. The uncertainty of the NLO pQCD calculations is shown in green dashed line, and the systematic uncertainty of experimental results is shown in filled boxes.}
\label{fig:tot_xsec}
\end{figure}

The detector acceptance$\times$efficiency corrected fraction of $J/\psi$ from $B$ meson decay with integrated $J/\psi$ $p_{T}$ ($p_{T}<5$ GeV/c) and $1.2<|y|<2.2$ rapidity coverage measured in 510 GeV $p$+$p$ collisions is shown in the left panel of Figure \ref{fig:bfrac_pp}. Reasonable agreement between this result and the Fixed-Order Next-To-Leading Logarithm and Color-Evaporation-Model (FONLL+CEM) \cite {Bedjidian2004gd, PhysRevLett.95.122001,vogt_dis} calculations has been achieved. Following the same analysis procedure, the $B \rightarrow J/\psi$ fraction is measured in 200 GeV $p$+$p$ collisions with the 2015 RHIC data. The right panel of Figure \ref{fig:bfrac_pp} shows the measured results in 200 GeV $p$+$p$ collisions with integrated $J/\psi$ $p_{T}$ in the forward and backward rapidity regions, which are also in good agreement with the derived rapidity dependent $B \rightarrow J/\psi$ fractions based on the FONLL \cite{FONLL} and the CEM \cite{CEM} calculations.

\section{Energy dependent $B \to J/\psi$ fraction and extrapolated $b\bar{b}$ cross section in $p$+$p$ collisions}
In order to understand the energy dependent $B$ hadron production, the PHENIX measured $B \rightarrow J/\psi$ fractions in 510 and 200 GeV $p$+$p$ collisions are compared with other Tevatron and LHC measurements \cite{Acosta:2004yw, Abelev:2012gx, Khachatryan:2010yr,lhcb8TeV, Aaij:2015rla} with integrated $J/\psi$ $p_{T}$ up to 5 GeV/c at different rapidities and higher center of mass energies. As shown in Figure \ref{fig:energy_btojpsi}, the center of mass energy dependent $B \rightarrow J/\psi$ fraction results indicate a smooth transition of $B$ production from the RHIC energies which are dominated by the gluon-gluon fusion process to the LHC energies which are dominated by the gluon splitting process. To further confirm this, the total $b\bar{b}$ cross sections in 200 and 510 GeV $p$+$p$ collisions are extracted from the forward $B \to J/\psi$ fraction results at PHENIX with the measured inclusive $J/\psi$ cross section \cite{Adare:2011vq}, the branching ratio of $B$ meson decay to $J/\psi$ and the phase space scaling factor calculated by the FONLL. The preliminary results of the extracted $b\bar{b}$ cross sections in $p$+$p$ collisions, as shown in Figure \ref{fig:tot_xsec}, are in reasonable agreement with the NLO pQCD calculations \cite{Vogt2008}. The PHENIX extracted $b\bar{b}$ cross sections and other global measurements follow the NLO pQCD predicted energy dependence of the bottom production.

\section{Summary and Outlook}
First measurements of $J/\psi$ from $B$ meson decay in forward/backward rapidities with integrated $p_{T}$ starting from zero are achieved in both 510 and 200 GeV $p$+$p$ collisions at PHENIX. A smooth transition is found for the $B \rightarrow J/\psi$ fractions with integrated $J/\psi$ $p_{T}$ up to 5 GeV$/c$ measured in $p$+$p$ or $p$+$\bar{p}$ collisions from 0.2 to 13 TeV. The NLO pQCD calculated $b\bar{b}$ cross sections are consistent with the extracted values based on the $B \rightarrow J/\psi$ fraction measurements at PHENIX in 200 and 510 GeV $p$+$p$ collisions. Global data on the bottom cross section follow the energy dependence predicted by the NLO pQCD calculations.

Large data sets in $p$+$p$ and $p$+Au collisions collected by PHENIX provide opportunities to further study the open heavy flavor production in forward/backward rapidities. This allows us to study the $p_{T}$ dependent $B \rightarrow J/\psi$ yields and explore the cold nuclear matter effects on the bottom production. The method used for the $B$ meson decay to $J/\psi$ fraction analysis through analyzing the muon $DCA_{R}$ can be extended to the study of $B$ and $D$ meson semi-leptonic decays to muons in the forward/backward rapidities to study the charm and bottom production in the low $p_{T}$ region.


\begin{thebibliography}{99}
\bibitem{Norrbin:2000zc}
Norrbin E and Sjostrand T 2000 {\em Eur. Phys. J. C\/} {\bf 17} 137
  (\textit{Preprint} arXiv: hep-ph/0005110)

\bibitem{ppg197}
Aidala C {\em et~al.\/} (PHENIX Collaboration) 2017 {\em Phys. Rev. D\/} {\bf          
  95} 092002 (\textit{Preprint} arXiv:1701.01342)

\bibitem{ppg199}
Aidala C {\em et~al.\/} 2017  (\textit{Preprint} arXiv:1702.01085)

\bibitem{Bedjidian2004gd}
Bedjidian M {\em et~al.\/} 2004 {Hard probes in heavy ion collisions at the
  LHC: Heavy flavor physics} (\textit{Preprint} arXiv: hep-ph/0311048)

\bibitem{PhysRevLett.95.122001}
Cacciari M, Nason P and Vogt R 2005 {\em Phys. Rev. Lett.\/} {\bf 95}(12)
  122001

\bibitem{vogt_dis}
Nelson R~E, Vogt R and Frawley A~D 2013 {\em Phys. Rev. C\/} {\bf 87} 014908
  (\textit{Preprint} arXiv: 1210.4610)

\bibitem{FONLL}
M~Cacciari M~G and Nason P 1998 {\em JHEP\/} {\bf 9805} 007 (\textit{Preprint}
  arXiv: hep-ph/9803400)

\bibitem{CEM}
Frawley A~D, Ullrich T and Vogt R 2008 {\em Phys. Rept.\/} {\bf 462} 125--175
  (\textit{Preprint} arXiv: 0806.1013)

\bibitem{Acosta:2004yw}
Acosta D {\em et~al.\/} 2005 {\em Phys. Rev. D\/} {\bf 71} 032001

\bibitem{Abelev:2012gx}
Abelev B {\em et~al.\/} 2012 {\em JHEP\/} {\bf 11} 065

\bibitem{Khachatryan:2010yr}
Khachatryan V {\em et~al.\/} 2011 {\em Eur. Phys. J. C\/} {\bf 71} 1575

\bibitem{lhcb8TeV}
Aaij R {\em et~al.\/} (LHCb Collaboration) 2013 {\em JHEP\/} {\bf 06} 064
  (\textit{Preprint} arXiv: 1304.6977)

\bibitem{Aaij:2015rla}
Aaij R {\em et~al.\/} 2015 {\em JHEP\/} {\bf 10} 172

\bibitem{Adare:2011vq}
Adare A {\em et~al.\/} (PHENIX Collaboration) 2012 {\em Phys. Rev. D\/} {\bf           
  85} 092004 (\textit{Preprint} arXiv: 1105.1966)

\bibitem{Vogt2008}
Vogt R 2008 {\em The European Physical Journal Special Topics\/} {\bf 155}
  213--222

\end{thebibliography}

\end{document}